\documentclass[preprint,amsmath,amssymb,showpacs]{revtex4}
\begin{document}
%
%
%
%
\title{Simplicial Gravity and Strings}
\author{John Swain}
\address{Department of Physics, Northeastern University, Boston, MA 02115, USA\\
email: john.swain@cern.ch
}
\date{October 22, 2011}


\vspace*{-0.7cm}
\begin{abstract}
String theory, as a theory containing quantum gravity, is
usually thought to require more dimensions of spacetime
than the usual 3+1. Here I 
argue on physical grounds that 
needing extra dimensions for strings may well be an
artefact of forcing a fixed flat background space. I also show that discrete
simplicial approaches to gravity in 3+1 dimensions have natural
string-like degrees of freedom which are inextricably tied to the dynamical space
in which they evolve. In other words, if simplicial approaches to 3+1 dimensional
quantum gravity do indeed give consistent theories, they may essentially
contain consistent background-independent string theories.
\end{abstract}


\maketitle
\clearpage

\section{Introduction}

String theories are well known to require unphysical (not 3 space and 1 time)
numbers of dimensions in which to propagate. Here I briefly review
why this might reasonably be expected on physical grounds. 
I then show how approaches to simplicial quantum gravity could 
contain consistent theories of 1-dimensional string-like objects.

In other words, it might be that rather than getting gravity from strings, we could
get strings from gravity.

\section{Why should closed string theories contain general relativity?}

There are at least two reasons why one might expect string theory to 
contain gravity. One is that closed strings have modes of vibration which can be
interpreted as spin-2 massless particles, and the only consistent coupling
of such particles seems to be linearized general relativity \cite{Polchinski}.
This is an essentially perturbative result, with the hope that the full nonlinear
theory can be correctly represented by infinite sums of graviton exchanges
(although it does take some strain to imagine getting a black hole out of
lots of spin-2 gravitons in flat space)!

The other is to view closed strings as maps from loops with the topology
of the circle into some background spacetime manifold $M$. The theory
is then really one of the infinite dimensional space of loops on $M$. This
loop space is a very complicated object but in the limit of shrinking the loops
to points one sees just the geometry of points of $M$ and recovers the usual notions of differential geometry. 
This would lead
one to expect that the geometry of infinite dimensional loop space would naturally
contain the finite dimensional geometry of space, and indeed Bowick and Rajeev\cite{Bowick-Rajeev}
were able to argue eloquently that string theory could be seen as the K\"{a}hler geometry of
loop space. 

\section{The unphysical spacetime dimensionalities of string theory}

Let us quickly review why string theory as usually formulated leads
to the need for extra spacetime dimensions.

The usual argument is to try to quantize strings propagating in a flat fixed background
spacetime ${\mathbb{R}}^{n+1}$ with $n$ space dimensions and one time.
Consistent quantization  \cite{Polchinski} then requires that the spacetime have certain
critical dimensions: 25+1 if the string
is assumed to be bosonic, and 9+1 if the string is given fermionic degrees
of freedom and required to be supersymmetric. 

From the loop space point of view, requiring an appropriately defined Ricci tensor
for loop space to vanish as an analog of Einstein's equations leads, for strings in flat
spacetime, to require the same unphysical dimensions\cite{Bowick-Rajeev}.

The most common response to these extra dimensions has been to find ways to 
argue that they are compactified at a scale so small as to render them effectively
inaccessible to us, or re-interpret them as additional fields in the theory\cite{Polchinski}.
I would like to argue, however, that there is a very
good physical reason to expect nonsensical predictions for the dimension
of a flat nondynamic spacetime in a theory that one hopes would reproduce general relativity.

\section{Why should string theory require unphysical numbers of spacetime dimensions?}

The key point is that Einstein's equations tell us that spacetime is dynamical
and is curved by the presence of matter and energy. If a string carries energy
it's hard to see why it would make sense then, given that one wants a theory
that will reproduce the results of general relativity, to expect consistent solutions
with zero spacetime curvature where the strings are.

Indeed, simply relaxing the requirement that the background be flat, as is done
in studies of strings propagating on group manifolds\cite{strings-on-groups}, one
can find lower critical dimensions, getting closer to our own 3+1. Even then,
the assumption of a completely homogeneous space with no localized curvature
where a string is, seems bizarre from a physical point of view.

People working in loop quantum gravity \cite{LQG} have long argued for the importance
of background independence and the development of a relational theory -- a view that
I personally would agree with -- but I think the argument given here of what's physically
inconsistent 
with the usual approach to string theory is quite robust and independent
of any particular philosophical viewpoint: {\em The conditions under which unphysical dimensions
are derived are also those which would require the theory to behave in a way that is
directly in conflict with the physical content of Einstein's equations.} \footnote{Quite amazingly,
the extra dimensions are often compactified as flat tori, with zero curvature,
or Ricci-flat ones selected essentially by hand in the hopes of finding some sort of
phenomenology that has some sort of resemblance to the real world. Disallowing
dynamics, or even any Ricci curvature at all (!) for the spaces of the hidden dimensions is as strange as 
doing it for the big ones, from the point of view of general relativity.}

Might string theory be viable if one did not insist on the physically
unreasonable ansatz of a flat nondynamical background (perhaps to be relaxed later
by some sort of back-reaction)? Perhaps, in a sense, string
theory, in a sense, back-reacts to having natural degrees of freedom frozen
by requiring extra dimensions.

\section{Strings in Simplicial Quantum Gravity}

One way of thinking about a discrete spacetime
is to imagine that space (or spacetime) is made of
piecewise flat simplices which fit
together to approximate a smooth geometry (for
an excellent review, see \cite{discrete}).
Examples include Regge
calculus\cite{Regge},
in which the dynamical degrees of freedom are taken to be
distances between the points that define the
simplices, and dynamical triangulation\cite{DT} in which those
distances are kept fixed but simplices are added and subtracted
as needed.  Such piecewise linear approximations can be used as
numerical approximations for classical general relativity, or
used to construct geometries to go into a path integral (or other)
types of quantization. 

Quite remarkably, simplicial spaces also arise naturally in loop quantum gravity\cite{LQG}, effectively
as duals to spin networks -- one dimensional graphs with edges labelled by $SU(2)$ representations,
and with interwiners at the vertices. The edges can be interpreted as quanta of area
and the vertices as quanta of space, and a quite direct connection can be made with a
picture of space as built up from polyhedra\cite{polyhedra} glued together. For earlier work
connecting spin networks and quantum gravity, see also reference \cite{Hasslacher}.

The point I want to make here is that simplical spaces, however they arise, naturally contain things one
can interpret as lower-dimensional objects which are intrinsically coupled
to a dynamical background. The idea is easy to understand:
In two dimensions
one can approximate a 2-manifold with flat triangles, joined in such a way 
that the curvature is only a points where the sums of the angles fail to add to $2\pi$.
In three dimensions the curvature is similarly distributional, but now along 1-dimensional
lines. With the usual association in general relativity of curvature and matter, one
could think of these as analogs of string-theoretic strings moving in a background
which is flat everywhere else. I propose that one calls any 1-dimensional distributional
curvature a ``simplicial string'' or ``S-string''.

It is interesting to note that the appearance of 1-dimensional objects is actually implied
by starting with a 3+1 dimensional spacetime, so there is a certain naturalness to the
idea. If one assumed a D+1 dimensional simplicial spacetime one would have been
led to distribution curvature on D-2 dimensional subspaces - sheets, or volumes, or
hypervolumes instead of 1-dimensional singular objects.

It is interesting to note the striking degree of similarity between the strings of the usual string theories
and S-strings, as well as their differences. Both are 1-dimensional objects in a space which is  flat
where the strings are not.
However, for regular strings, space is still flat where the strings {\em are}, while the space where
S-strings are, is, by definition, where curvature is localized! The space in which string
theories place the strings must be selected by hand and is static. In contrast, S-strings
are intrinsically part of a dynamical space and inseparable from it.

S-strings need not be closed, and indeed could form complex and branching
networks, dynamically evolving and changing in topology as the associated simplicial
3-space evolves. They thus represent a much richer and more complicated structure than
the simple closed loops in flat spacetime of string theory. The implications of this in terms of
what else might be hidden in simplicial quantum gravity have yet to be investigated, but there is
clearly much richness and unexplored structure present.

Note that while string theory postulates 1-dimensional strings and then tries to 
derive the dimensionality of space, here we take the physical dimensionality of
space as given and deduce, with the simplicial assumption\footnote{Which, as noted earlier,
arises naturally in loop quantum gravity.}, the existence of 1-dimensional
objects!

Einstein had lamented the form of the equation $R_{\mu\nu}=8\pi G T_{\mu\nu}$ as ugly, with the
left side completely geometrical and the right side essentially put in by hand. Here we see the 
possibility of a picture in which geometry is primary, with stringy matter at the smallest scales being
an aspect of space itself -- string are where the curvature is concentrated. Similarly, the fact that the
highly nonlinear Einstein equations  allow one to determine how matter moves in a gravitational
field (the geodesic equation) so that ``matter tells space how to curve and space tells matter how to
move'' \cite{MTW} finds an analog here in that S-strings are both
part of the geometry of space {\em and} what one might want to think of as matter, with their evolutions 
inextricably linked together.

\section{Conclusions}

I have argued that the appearance of unphysical spacetime dimensions found in theories of strings
which are supposed to contain gravity might very well be expected on physical grounds, and not
necessarily a reason to rule out fundamental 1-dimensional objects.

String theory over the last several years has been realized to be
a theory containing a variety of extended objects, perhaps most notably D-branes on which
strings can end\cite{D-branes}.  In a similar fashion, I have argued that any approach to simplicial quantum gravity
contains 1-dimensional degrees of freedom -- S-strings -- which appear naturally together with the geometry and
as an intrinsic part of it. S-strings, though genuinely one-dimensional objects which live in a
3-dimensional space, are not put in ``by hand'', but, rather, emerge naturally. They must co-evolve
with space and indeed have no meaning independent of it. There is no need to make assumptions about
some pre-existing background spacetime, much less that it be flat. If simplicial quantum gravity
is indeed present in a consistent theory of  quantum spacetime, it would effectively unify ``space'' and ``matter'' degrees of
freedom in a completely natural and relational way.

It is also interesting to see that apparently very different approaches to quantum gravity -- simplicial
quantum gravity, loop quantum gravity, and string theory - 
might have more points in the common than is usually recognized.

\section{Acknowledgements}

This work was supported in part by the US National Science Foundation
and grant NSF0855388.

\end{document}